\documentclass[preprint,amsmath,amssymb,superscriptaddress,showpacs]{revtex4}
\usepackage{graphicx}
\begin{document}
\renewcommand{\thesection}{\arabic{section}}
\renewcommand{\thesubsection}{\arabic{section}.\arabic{subsection}}
\renewcommand{\thefigure}{\arabic{figure}}
\baselineskip=0.7cm
\title{Quasiparticle properties of graphene in the presence of disorder}
\author{A. Qaiumzadeh}
\affiliation{Institute for Advanced Studies in Basic Sciences (IASBS), Zanjan,
P. O. Box 45195-1159, Iran}
\affiliation{School of Physics, Institute for Studies in
Theoretical Physics and Mathematics, 19395-5531 Tehran, Iran}
\author{N. Arabchi}
\affiliation{Department of Physics, Amir Kabir University of Technology, P. O. Box 158754413, Tehran, Iran}
\author{R. Asgari~\footnote{Corresponding author: Tel: +98 21 22280692; fax: +98 21 22280415.\\ E-mail address: asgari@theory.ipm.ac.ir, }}
\affiliation{School of Physics, Institute for Studies in
Theoretical Physics and Mathematics, 19395-5531 Tehran, Iran}

\begin{abstract}
We calculate the quasiparticle properties of chiral two-dimensional Dirac electrons in graphene
within the Landau Fermi Liquid scheme based on $GW$ approximation in the presence of disorder.
Disorder effects due to charged impurity scattering plays a crucial role in density dependence of quasiparticle quantities. Mode-coupling approach to scattering rate and self-energy in $GW$ approximation for quasiparticle renormalized Fermi velocity and spin-antisymmetric Landau Fermi parameter incorporating the many-body interactions and the disorder effects show reduction of these quantities by $5-15$
percent at available experimental charge carrier density region.
\end{abstract}
\maketitle

\section{Introduction}

Graphene, a single atomic layer of graphite, has been
successfully produced in experiment~\cite{novoselov}, which has
resulted in intensive investigations on graphene-based
structures because of fundamental physics interests and
promising applications~\cite{geim}. There are significant efforts to grow
graphene epitaxially~\cite{berger} by thermal decomposition of Silicon Carbide
(SiC), or by vapor deposition of hydrocarbons on catalytic metallic
surfaces which could later be etched away leaving graphene on an
insulating substrate.

The renormalized Fermi liquid
parameters for both undoped and doped graphene have been firstly
calculated by Das Sarma {\it et al}~\cite{sarma} focusing
on the role of intraband density functions and on analytic
weak-interaction expansions and they found
that for doped graphene renormalization factor, and effective Fermi velocity
show no deviation from the usual Fermi liquid behavior,
and the Fermi liquid description is robust. On the other
hand, with precisely zero doping, intrinsic graphene exhibits
a quasiparticle lifetime linear in the excitation energy and a
zero renormalization factor, indicating that the Fermi liquid
description is marginal at the Dirac point. Next, the inelastic carrier
lifetime in graphene~\cite{hwang1} and many-body electron-electron
quasiparticle properties in graphene have been correctly studied.~\cite{polini}

The quasiparticle energies and band gaps of graphene nanoribbons has
been calculated using a first-principles Green's function approach
within the $GW$ approximation~\cite{yang}. The electron self-energy,
spectral function, and the band velocity renormalization due to
phonon-mediated electron-electron interaction, has been calculated~\cite{tse}
and found that phonon-mediated electron-electron coupling has
a large effect on the graphene band structure renormalization
and a reduction of the band velocity by $10-20$ percent at
the experimental doping level.

Other unusual many-body features of the single-atom-thick layer of carbon
are the effects of electron-electron interactions on the ground-state
properties~\cite{yafis}, the plasmon behavior and essentially angle-resolved
photoemission spectroscopy (ARPES) ~\cite{polini2} that follow from chiral
band states and energy-momentum linear dispersion relation.

The low energy quasiparticle excitations
in graphene are linearly dispersing, described by Dirac
cones at the edges of the first Brillouin zone. The Linear energy-momentum
dispersion have been confirmed by recent observations~\cite{novoselov1}. The slope of this linearity corresponds the Fermi velocity of chiral Dirac electrons in graphene and it plays essential role in the Landau Fermi liquid theory and has a direct connection to experimental measurement. ARPES
is a powerful probe of spectral function, in two-dimensional crystals
because it achieves momentum ${\bm k}$ resolution~\cite{ARPES_RMP}.
Recent experiments~\cite{rotenberg,lanzara, zhou} have reported ARPES spectra for
single-layer graphene samples prepared by graphitizing the surface of SiC~\cite{deHeerSSC}. Zhou {\it et al}\cite{zhou} studied of the electronic structure of single layer epitaxial graphene with the main focus on the $\pi$ bands by ARPES. The dispersion of the $\pi$ bands
shows overall agreement with conical dispersions of Dirac
fermions. However, clear deviations from the conical dispersion
were observed near the Dirac point. They examined the two possible explanations for the experimental data,  many-body interactions and the opening of a gap at the
Dirac point due to breaking of sublattice symmetry.
Gr\"{u}neis {\it et al}\cite{grüneis} have performed ARPES of graphite
and compared the measured quasiparticle dispersion to {\it ab initio}
calculations. They have found that the band dispersions are
better described by the GW approximation, however around the K point that agreement becomes poorer and
many-body approximation schemes beyond GW may improve result. From these experiments view, the renormalized velocity can directly be measured from the quasiparticle dispersion relation at long wavelength region near to the Dirac point.

All the mentioned theoretical efforts calculating the
quasiparticle quantities have been carried out for clean graphene.
Since disorder is unavoidable in any material, there has been
great interest in trying to understand how disorder affects the
physics of electrons in material science specially here in
graphene and its quasiparticle properties.

Our aim in this work is to study the quasiparticle properties in
the presence of electron-impurity and electron-electron
interactions within the Landau Fermi Liquid theory where the concept of the model is applicable~\cite{Giuliani_and_Vignale} to impurity systems and inhomogeneous
electron liquids. It has been shown~\cite{fradkin} that the transport
of undoped graphene cannot
be described within the standard Boltzmann approach of
metals, because the Dirac-like electronic excitations have infinite
Compton wavelength violating the assumptions for the validity of Boltzmann
transport. Accordingly the proper way to compute the transport and quasiparticle properties are treating
the impurities in a self-consistent way. For this purpose, we use the self-consistent theory
of G{\"o}tze \cite{gotze} to calculate the scattering
rate, self-energy and the quasiparticle properties of the system at
the level of random-phase approximation (RPA) including disorder effects.
Implementing the impurity effects in our calculation, is in the same way of our earlier work on both chiral two-dimensional electron systems (C2DES) ~\cite{asgari_im} where we have calculated the ground-state properties and thermodynamics of graphene and
conventional two-dimensional electron gas (2DEG).\cite{asgari}

In this work, we consider the charged impurity and the
surface-roughness potentials which are always present in currently available graphene samples~\cite{meyer, ishigami}. It has been
demonstrated that a short-range scattering potential is irrelevant
for electronic properties of graphene~\cite{katsnelson} and impurity scattering in graphene is dominated by charged impurities and it is therefore strongly dependent on the carrier charge density~\cite{nomura}. We
have used the same method ~\cite{asgari, asgari3} to investigate
some properties of the conventional 2DEG. In this paper, we find that there are significant differences between quasiparticle properties of C2DES and conventional 2DEG due to disorder effects.

The rest of this paper is organized as follows. In Sec.\,II, we
introduce the models for self-consistent calculation of impurities
effect. We then outline the calculation of self-energy and quasiparticle quantities.
Section III contains our numerical calculations. We conclude in Sec.\,IV with a brief summary.

\section{Theoretical Model}

We consider a system of C2DES interacting via the
Coulomb potential $e^2/\epsilon r$ and its Fourier transform
$v_q=2\pi e^2/(\epsilon q)$
where $\epsilon$ is the background dielectric constant at zero temperature in the absence of electron-phonon and spin-orbit interactions.
The massless Dirac band Hamiltonian of graphene can be written as
${\cal H}= v \tau \left(\sigma_1 \, k_1 + \sigma_2 \, k_2 \right)$,
where $\tau = \pm 1$ for the inequivalent $K$ and $K'$ valleys at
which $\pi$ and $\pi^*$ bands
touch, $k_i$ is an envelope function momentum operator, $v$ is the
Fermi velocity and $\sigma_i$ is a Pauli matrix which acts on the
sublattice pseudospin degree-of-freedom.

Our results for C2DES are based on the
GW-RPA in which the self-energy is expanded to first order in the dynamically screened
Coulomb interaction $W$:
\begin{eqnarray}\label{eq:sigma_rpa}
\Sigma_s({\bf k},i\omega_n, \Gamma)&=&-\frac{1}{\beta}\sum_{s'}
\int \frac{d^2{\bf q}}{(2\pi)^2}\sum_{m=-\infty}^{+\infty}
W({\bf q},i\Omega_m, \Gamma)\nonumber\\
&\times&\left[\frac{1+s s'\cos{(\theta_{{\bf k},{\bf k}+{\bf q}})}}{2}\right]G^{(0)}_{s'}({\bf k}+{\bf q},i\omega_n+i\Omega_m)\,,
\end{eqnarray}
where $s=+$ for electron-doped systems and $s=-$ for
hole-doped systems, $\beta=1/(k_{\rm B} T)$, $\omega_n=(2n+1)\pi/\beta$ is a fermionic Matsubara frequency, $G^{(0)}_{s'}({\bf k}, i\omega)$ is noninteracting Green's function and the effective electron-electron interaction in $GW$ approximation is given by
\begin{equation}\label{eq:ex+corr}
W({\bf q},i\Omega, \Gamma)=v_q+v^2_q\frac{\chi^{(0)} ({\bf
q},i\Omega,\mu, \Gamma)} {1-v_q\chi^{(0)}({\bf q},i\Omega,\mu,
\Gamma)}~,
\end{equation}

where $\chi^{(0)}({\bf q},i\Omega,\mu,\Gamma)$ is the dynamical polarizability of noninteracting C2DES in which chemical
potential and the strength of damping are represented by $\mu$ and $\Gamma$, respectively. In the C2DES, dielectric-function contributions from intraband and interband excitations are subtly interrelated.  The two contributions must be included on an equal footing in order to describe C2DES physics correctly.

A central quantity in the theoretical formulation of the many-body
effects in Dirac fermions is the dynamical polarizability~\cite{shung, yafis, others} tensor
$\chi^{(0)}({\bf q},i\Omega,\mu\neq 0)$. This is defined through the one-body noninteracting
Green's functions.\cite{gonzalez_1994} To describe the properties of C2DES we define a dimensionless coupling constant $\alpha_{gr}=g{e^2/\upsilon \epsilon \hbar}$ where $g=g_vg_s=4$ is
the valley and spin degeneracy.

The effect of disorder is to dampen the charge-density fluctuations
and results to modify the dynamical polarizability tensor.
Within the relaxation time approximation the modified
$\chi^{(0)}({\bf q},i\Omega,\mu,\Gamma)$ is given by~\cite{mermin}
\begin{equation}
\chi^{(0)}({\bf q},i\Omega,\mu, \Gamma)= \frac{\chi^{(0)}({\bf
q},i\Omega+i\Gamma,\mu)}{1- \frac{\Gamma}{\Omega+\Gamma}\left[1-
\frac{\chi^{(0)}({\bf q},i\Omega+i\Gamma,\mu)}{\chi^{(0)}({\bf q})}
\right]}~.
\end{equation}
We
consider long-ranged charged impurity scattering and surface
roughness as the main sources of disorder. The latter mechanism
also known as ripples comes either from thermal fluctuations or
interaction with the substrate.\cite{nima} The disorder averaged
surface roughness (ripples) potential is modeled as $
\langle
|U_{surf}(q)|^2\rangle = \pi\Delta^2h^2 (2\pi e^2 n/\epsilon)^2
e^{-q^2\Delta^2/4} $
where $h$ and $\Delta$  are parameters
describing fluctuations in the height and width, respectively. We
can use the experimental results of Meyer {\it al}.\cite{meyer}
who estimate $\Delta\sim 10$\,nm and $h\sim 0.5$\,nm. It is
important to point out that there are other models to take into
account the surface-roughness potential. The effect of bending of the
graphene sheet has been studied by Kim and Castro Neto~\cite{kim}.
This model has two main effects, firstly the decrease of the
distance between carbon atoms and secondly a rotation of the $p_z$
orbitals. Due to bending the electrons are subject to a potential
which depends on the structure of the graphene sheet.
Another possible model is described by
Katsnelson and Geim ~\cite{katsnelson} considering
the change of in-plane displacements and out-of-plane displacements due to the
local curvature of a graphene sheet. Consequently, the change
of the atomic displacements results to change in nearest-neighbour
hopping parameters which is equivalent to the appearance of a random
gauge field described by a vector potential. These different models
need to be implemented in our scheme and to be checked numerically
to assess their validity in comparison to the available measurements.

The charged disorder potential is taken to be $
\langle |U_{imp}(q)|^2\rangle=n_i v_q^2 e^{-2qd}$
in which $n_i$ is the
density of impurities and $d$ is the setback distance from the
graphene sheet.

We modify the mode-coupling approximation introduced by
G{\"o}tze\cite{gotze} for C2DES to express the total scattering rate
in terms of the screened disorder potentials
\begin{displaymath}
i\Gamma=-\frac{v_F k_F }{2 \hbar n A}\sum_{\bf q}
\left[\frac{<\mid U_{imp}(q)\mid^2>}{\varepsilon^2(\bf
q)}\right.
\end{displaymath}
\begin{equation}
\left. +\frac{\langle|U_{surf}(q)|^2\rangle}{\varepsilon^2(\bf
q)}\right]\frac{\varphi_0({\bf q},i\Gamma)}{1+i\Gamma\varphi_0({\bf q},i\Gamma)/\chi^{0}({\bf q})}~,
\end{equation}
where $A$ is the sample area, $\varepsilon({\bf q})=1-v_q \chi^{(0)}({\bf
q})$ is the static screening function and the relaxation function for electrons
scattering from disorder is given as
$\varphi_0({\bf q},i\Gamma)=
\left[\chi^{(0)}({\bf q},i\Gamma,\mu)-\chi^{(0)}({\bf q})\right]/i\Gamma$ .

Since the scattering rate $\Gamma$ depends on the relaxation function
$\varphi_0({\bf q},i\Gamma)$, which itself is determined by the disorder
included response function, the above equation needs to be
solved self-consistently to yield eventually the
scattering rate as a function of the coupling constant.

\begin{figure}[ht]
\begin{center}
\tabcolsep=0 cm
\includegraphics[width=0.75\linewidth]{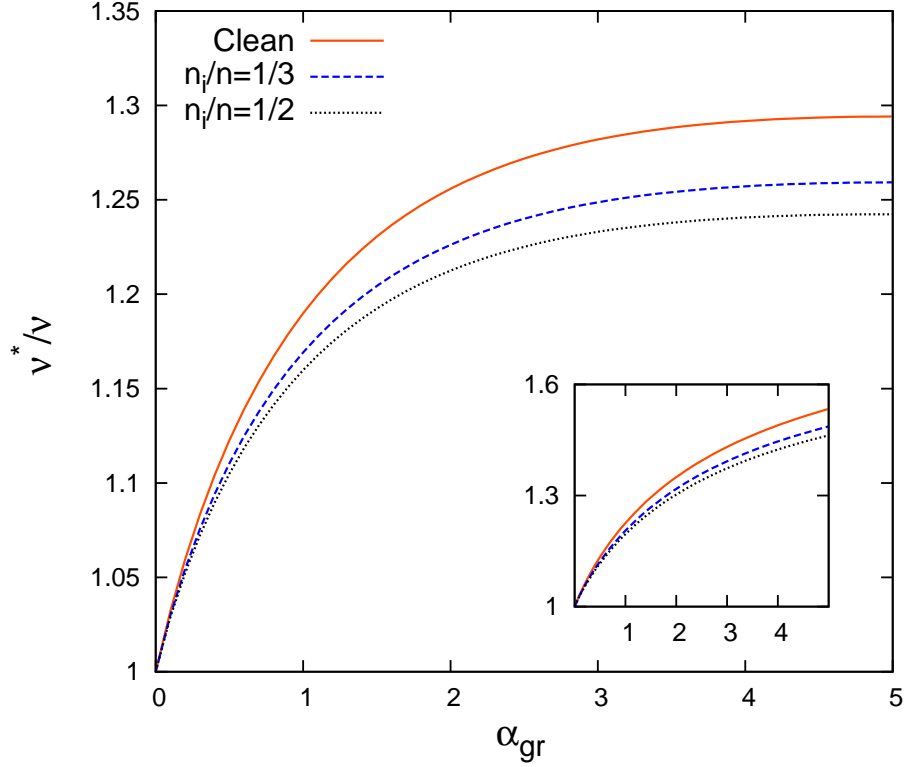}
\caption{(Color online) The renormalized Fermi velocity scaled by Fermi velocity of a noninteracting clean system as a function of the coupling constant $\alpha_{gr}$ for cut-off value $\Lambda=k_c/k_F=100$ for different charged impurity densities value. In the inset, the renormalized Fermi velocity scaled by noninteracting one as a function of the coupling constant $\alpha_{gr}$ for cut-off value $\Lambda=100$ calculated within on-shell approximation. }
\end{center}
\end{figure}

In Eq.~(\ref{eq:sigma_rpa}) the sum runs over all the bosonic Matsubara frequencies $\Omega_m=2m\pi/\beta$. The factor
in square brackets in Eq.~(\ref{eq:sigma_rpa}), which depends on the angle $\theta_{{\bf k},{\bf k}+{\bf q}}$ between ${\bf k}$ and ${\bf k}+{\bf q}$, captures
the dependence of Coulomb scattering on the relative chirality $s s'$ of the interacting electrons.
The Green's function $ G^{(0)}_{s}({\bf k},i\omega) = 1/[i\omega - \xi_s({\bf k})]$ describes the free propagation of
states with wavevector ${\bf k}$, Dirac energy $ \xi_s({\bf k})=sv k-\mu$ (relative to the chemical potential) and chirality $s=\pm$. The first and second terms in Eq.~(\ref{eq:ex+corr}) are responsible respectively for the exchange interaction with the
occupied Fermi sea (including the negative energy component),
and for the interaction with particle-hole and collective virtual fluctuations.
After continuation from imaginary to real frequencies, $i \omega \to
\omega + i \eta$, the quasiparticle weight factor
$Z$ and the renormalized Fermi velocity can be expressed~\cite{Giuliani_and_Vignale} in terms of the wavevector and
frequency derivatives of the retarted self-energy $\Sigma^{\rm ret}_+({\bf k},\omega,\Gamma)$ evaluated at the Fermi surface ($k=k_{\rm F}$) and
at the quasiparticle pole $\omega=\xi_+({\bf k})$:
\begin{equation}\label{eq:Z_def}
Z= \frac{1}{1-\left.\partial_{\omega} \Re e \Sigma^{\rm ret}_+({\bf k},\omega,\Gamma)\right|_{k=k_{\rm F},\omega=0}}\,,
\end{equation}
and
\begin{equation}\label{eq:v_star_dyson}
\frac{v^\star}{v}
=\frac{\displaystyle 1+(v)^{-1}\left.\partial_k \Re e \Sigma^{\rm ret}_+({\bf k},\omega,\Gamma)\right|_{k=k_{\rm F},\omega=0}}{Z^{-1}}\,.
\end{equation}
On the other hand,  we find that the quasiparticle renormalized Fermi velocity
within the on-shell approximation is given by $v^\star/v
=\displaystyle 1+(v)^{-1}\left.\partial_k \Re e \Sigma^{\rm ret}_+({\bf k},\omega,\Gamma)\right|_{k=k_{\rm F},\omega=0}+\left.
\partial_{\omega} \Re e \Sigma^{\rm ret}_+({\bf k},\omega,\Gamma)\right|_{k=k_{\rm F},\omega=0}$.
Evidently, the on-shell approximation is a valid expression for the quasiparticle renormalized Fermi velocity only in the weak coupling limit where $\alpha_{gr}$ becomes small enough. The physical reasons are discussed in Ref.~[\onlinecite{asgari2}] in the context of conventional 2DEG.

Following some standard manipulations~\cite{Giuliani_and_Vignale} the self-energy can be expressed in
a form convenient for numerical evaluation, as the sum of a contribution from the interaction of quasiparticles at the Fermi energy, the {\em residue} contribution $\Sigma^{\rm res}$, and a contribution from interactions with quasiparticles far from the Fermi energy and via both exchange and virtual fluctuations, the {\em line} contribution $\Sigma^{\rm line}$.
In the zero-temperature limit
\begin{eqnarray}\label{eq:residue_t_0}
\Sigma^{\rm res}_+({\bf k},\omega,\Gamma)&=&\sum_{s'}\int \frac{d^2 {\bf q}}{(2\pi)^2}
\frac{v_q}{\varepsilon({\bf q},\omega-\xi_{s'}({\bf k}+{\bf q}),\Gamma)}\left[\frac{1+s'\cos{(\theta_{{\bf k},{\bf k}+{\bf q}})}}{2}\right]\nonumber\\
&\times&\left[\Theta(\omega-\xi_{s'}({\bf k}+{\bf q}))-\Theta(-\xi_{s'}({\bf k}+{\bf q}))\right]
\end{eqnarray}
and
\begin{equation}\label{eq:line_t_0_better}
\Sigma^{\rm line}_+({\bf k},\omega,\Gamma)=-\sum_{s'}
\int \frac{d^2 {\bf q}}{(2\pi)^2}v_q\left[\frac{1+s'\cos{(\theta_{{\bf k},{\bf k}+{\bf q}})}}{2}\right]\int_{-\infty}^{+\infty}\frac{d\Omega}{2\pi}
\frac{1}{\varepsilon({\bf q},i\Omega,\Gamma)}\frac{\omega-\xi_{s'}({\bf k}+{\bf q})}{[\omega-\xi_{s'}({\bf k}+{\bf q})]^2+\Omega^2}\,,
\end{equation}
where $\varepsilon({\bf q},i\Omega,\Gamma)=1-v_q \chi^{(0)}({\bf
q},i\Omega,\mu,\Gamma)$ is the dynamic dielectric function in RPA level.
Note that at the Fermi energy $\partial_{k} \Sigma^{\rm res}_+({\bf k},\omega,\Gamma)$ vanishes,
and $\partial_{\omega} \Sigma^{\rm res}_+({\bf k},\omega,\Gamma)$ involves an integral over interactions on the Fermi surface that are statically screened.  These expressions differ from the corresponding 2DEG expressions because of the relative chirality dependence of the Coulomb matrix elements, because of the
linear dispersion of the bare quasiparticle energies, and most importantly because of the
fast short-wavelength density fluctuations produced by the interband contribution to
$\chi^{(0)}({\bf q},i\Omega, \Gamma)$. we can introduce an ultraviolet cutoff
for the wave vector integrals $k_c=\Lambda k_F$ which is the order
of the inverse lattice spacing and $\Lambda$ is dimensionless
quantity. Fermi momentum is related to density as given by
$k_F=(4\pi n/g)^{1/2}$.

\begin{figure}[ht]
\begin{center}
\includegraphics[width=0.75\linewidth]{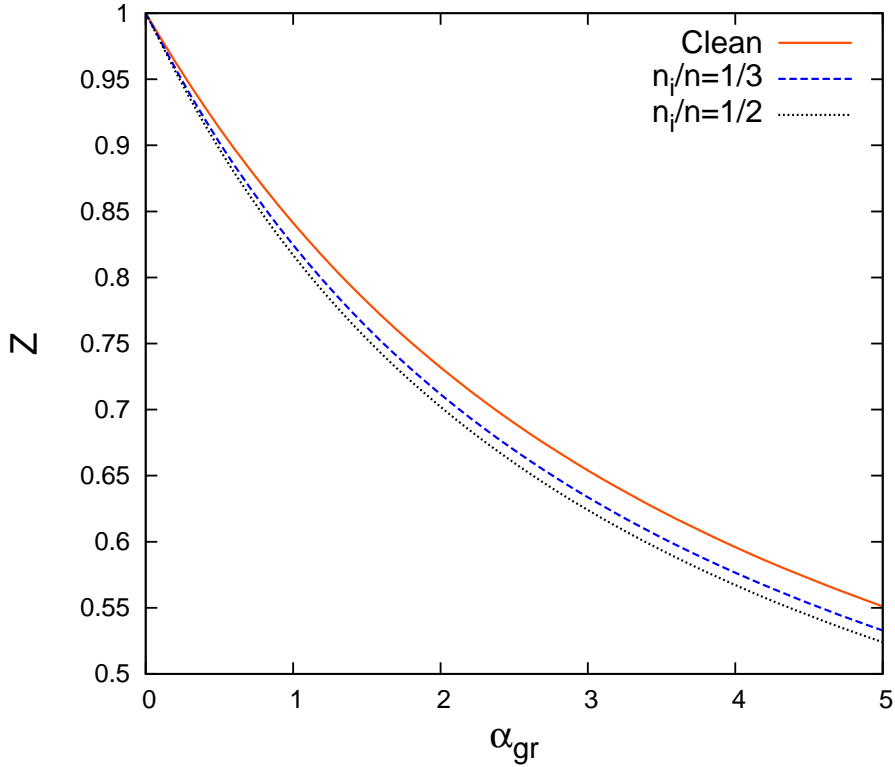}
\caption{(Color online) The quasiparticle weight factor $Z$
as a function of the coupling constant
$\alpha_{gr}$ for cut-off value $\Lambda=k_c/k_F=100$
for different charged impurity densities.}
\end{center}
\end{figure}

\begin{figure}[ht]
\begin{center}
\includegraphics[width=0.75\linewidth]{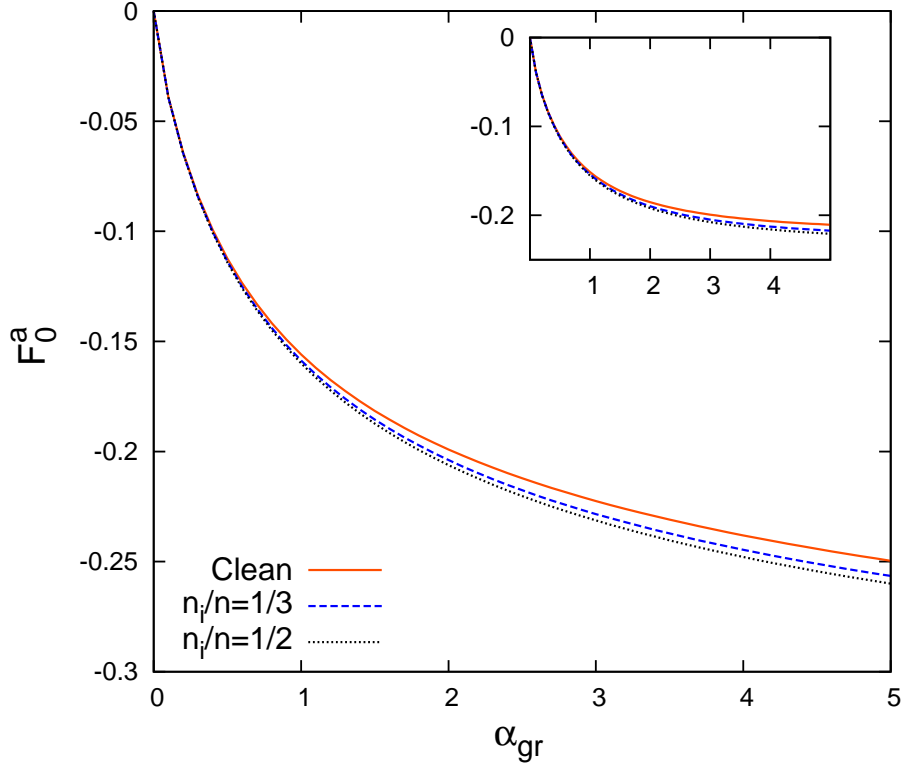}
\caption{(Color online) The spin-antisymmetric Landau Fermi parameter
$F^a_0$ as a function of the coupling constant
$\alpha_{gr}$ for cut-off value $\Lambda=k_c/k_F=100$.
In the insert, the spin-antisymmetric Landau Fermi parameter
$F^a_0$ as a function of the coupling constant
$\alpha_{gr}$ calculated within the on-shell renormalized Fermi velocity approximation.}
\end{center}
\end{figure}

In order to have in connection with experiments, we would like to calculate the spin-antisymmetric Landau Fermi parameter, $F^{\rm a}_0$ which is related to the spin-susceptibility, in which $\chi_0/\chi_s=v^*(1+F^{\rm a}_0)/v$. For this purpose, we obtain the Landau interaction function $f^{\sigma, \sigma'}_{{\bf k}, {\bf k}'}$ are defined by the following functional derivative~\cite{Giuliani_and_Vignale}
\begin{equation}\label{eq:interaction_first}
f^{\sigma, \sigma'}_{{\bf k}, {\bf k}'}=\frac{\delta \Re e \Sigma^{\rm ret}_\sigma({\bf k},vk)}{\delta n^{(0)}_{+, \sigma', {\bf k}'}},
\end{equation}
where the occupation numbers $n^{(0)}_{+, \sigma, {\bf k}}$ in conduction band is defined as $\Theta({\bf k}_{F_{\sigma}}-{\bf k})$.
The spin-antisymmetric Landau Fermi interaction function $f_a(\varphi)$ is defined by $\frac{1}{2}(f^{\uparrow, \uparrow}_{{\bf k}, {\bf k}'}-f^{\uparrow, \downarrow}_{{\bf k}, {\bf k}'})$. After some straightforward algebraic calculations, $f_{\rm a}(\varphi)$ on the Fermi surface is given by
\begin{equation}
f_{\rm a}(\varphi)=
-\frac{1}{2}\left[\frac{1+\cos{(\theta_{{\bf k}'-{\bf k},{\bf k}'})}}{2}\right]W({\bf k}'-{\bf k},0)\,,
\end{equation}
where $\varphi$ being the angle between ${\bf k}$ and ${\bf k}'$ and $|{\bf k}| = |{\bf k}'| = k_F$.
Furthermore, $F^{\rm a}_0$ calculated using the following equation
\begin{equation}
F^{\rm a}_0=\frac{\hbar k_{\rm F}}{v^\star\pi}
\int_0^{2\pi}\frac{d\varphi}{2\pi}f_{\rm a}(\varphi)
\end{equation}

\section{Numerical results}

In this section we present our numerical calculations for quasiparticle
properties of graphene in the presence of impurities which are modeled as mentioned above. The quasiparticle properties are calculated by using
the theoretical models described above and the impurities dependence of quasiparticle quantities are calculated.
In all numerical calculations we consider $d=0.5$\,nm. Electron density
is taken to be $1\times 10^{12}$\,cm$^{-2}$ for Figs.\,1-3.

Increasing disorder (increasing $n_i$ or
decreasing $d$ for charge-disorder potential or increasing $h$ for
surface roughness potential) decrease the
$\chi^{(0)}(q,\Omega,\mu,\Gamma)$ as the scattering rate $\Gamma$
gets bigger. Thus, decreasing $\chi^{(0)}(q,\Omega,\mu,\Gamma)$
(or increasing correlation effects) results in a stronger disorder
potential. Despite $\Gamma$ increases with increasing
$\alpha_{gr}$, apparently it turns to a saturation limit and does
not diverge~\cite{asgari_im}. This behavior is quite different than what is seen in
conventional 2DEG ~\cite{asgari} when the many-body effects
influence the scattering rate through the local-field factor. In
the conventional 2DEG system, at a critical level of disorder this
nonlinear feedback causes $\Gamma$ to increase rapidly and diverge,
which is taken as an indication for the localization of carriers. However,
in graphene, our calculations show that the $\Gamma$ does not diverge
therefore impurities cannot localize carriers and we have a weakly localized
system in the presence of impurities compatible with experimental
observations~\cite{morozov}. We have found through our
calculations~\cite{asgari_im} that $\Gamma$ increases with increasing $n_i/n$ as a function of $\alpha_{gr}$.

The Fermi liquid phenomenology of Dirac electrons in
graphene~\cite{polini, polini2} and conventional
2DEG~\cite{asgari2} have the same structure, since both systems are
isotropic and have a single circular Fermi surface. The strength of
interaction effects in a conventional 2DEG increases with decreasing
carrier density. At low densities, the quasiparticle weight factor $Z$ is
small, the velocity is suppressed~\cite{asgari2}, the charge
compressibility changes sign from positive to negative\cite{asgari},
and the spin-susceptibility is strongly enhanced~\cite{asgari2}.
These effects emerge from an interplay between exchange interactions
and quantum fluctuations of charge and spin in the 2DEG.

In the C2DEG clean graphene, it was
shown~\cite{yafis,polini,polini2} that interaction effects also
become noticeable with decreasing density, although more slowly,
the quasiparticle weight $Z$ tends to larger values, that the
velocity is enhanced rather than suppressed, and that the
influence of interactions on the compressibility and the
spin-susceptibility changes sign. These qualitative differences
are due to exchange interactions between electrons near the Fermi
surface and electrons in the negative energy sea and to interband
contributions to Dirac electrons from charge and spin
fluctuations.

\begin{figure}[ht]
\begin{center}
\includegraphics[width=0.75\linewidth]{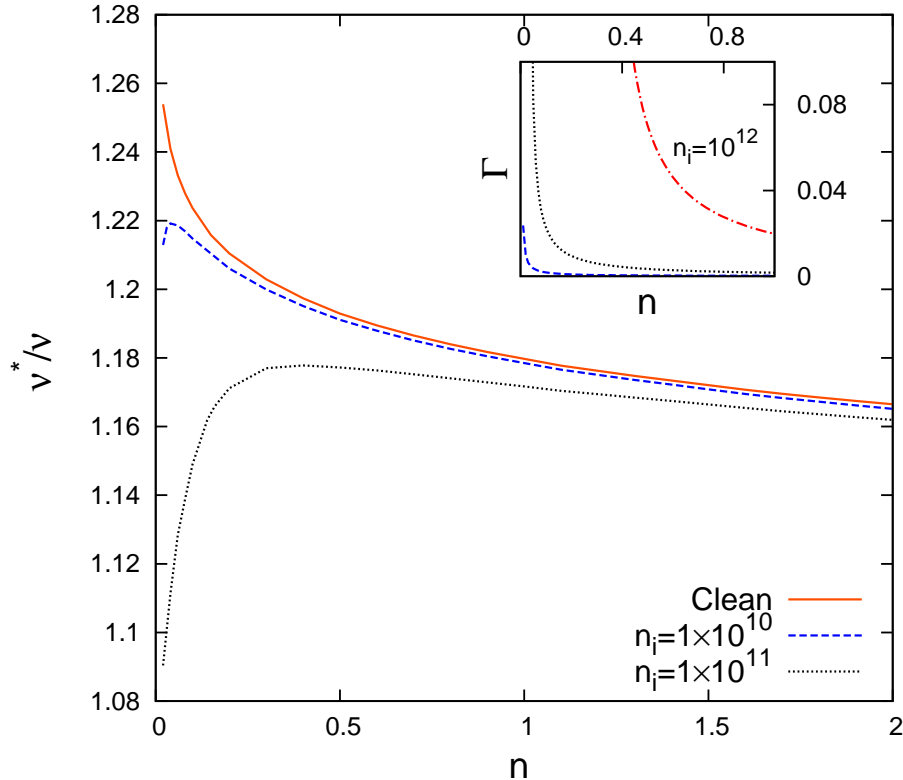}
\caption{(Color online) The renormalized Fermi velocity scaled by Fermi velocity of a noninteracting clean system, $v^*/v$ as a function of the electron charge density
$n$ in unite of $10^{12}~{\rm cm}^{-2}$ at $\alpha_{gr}=1$.
In the inset, the scattering rate $\Gamma$ as a function of electron charge density $n$ at $\alpha_{gr}=1$ for density impurities, $10^{10}, 10^{11}$ and $10^{12}~{\rm cm}^{-2}$, respectively.}
\end{center}
\end{figure}

We have calculated the renormalized Fermi velocity scaled by Fermi velocity of noninteracting clean system as a function of $\alpha_{gr}$ in the presence of disorder as shown in Figure 1. It is found
that the disorder effects become more appreciable at large
coupling constants, within the mode coupling approximation.
The renormalized Fermi velocity becomes smaller due to disorder effect about $10-15$ percent at available experimental $\alpha_{gr}$ value, namely smaller than 2 ( $\alpha_{gr}\approx 1$ or 2 is a typical value thought to apply to graphene sheets on the surface of a (${\rm Si/SiO}_2$ or ${\rm SiC}$) or  ${\rm SiO}_2$ substrate, respectively). Since the scattering rate $\Gamma$ tends to a
constant at large $\alpha_{gr}$, hence the effect of disorder for
$v^*/v$ is fixed at large $\alpha_{gr}$.
In the insert of figure~1, we have shown the results of renormalized Fermi velocity calculated within the on-shell approximation. As we pointed out the on-shell approximation is valid at small value of
$\alpha_{gr}$. The behavior of $v^*$ shows some novel physics, which is
qualitatively different from the physics known in the conventional
2DEG. In the conventional 2DEG, effective mass decreases by increasing impurity at small coupling constant, namely $r_s <1$  and then enhances by increasing impurity effects at large $r_s$ value.~\cite{asgari3}

Figure~2 shows the quasiparticle weight factor $Z$ as a function of
$\alpha_{gr}$. It appears that the disorder effects become more
appreciable at large coupling constant and decreases by
increasing the disorder effects. Significantly, the $\partial_\omega \Re e \Sigma^{ret}_+({\bf k},\omega,\Gamma)$ does not change by impurity, however $\partial_\omega \Re e \Sigma^{line}_+({\bf k},\omega,\Gamma)$ has negative values which
decreases by increasing impurity density.
Moreover, the $\partial_k \Re e\Sigma^{line}_+({\bf k},\omega,\Gamma)$ has positive values which
increases slightly by increasing impurity density. Accordingly, the quasiparticle weight factor $Z$ decreases by increasing the impurities effect. Importantly, this behavior is in contrast with 2DEG where $Z$ increases by increasing impurity effects.

Figure~3 shows the spin-antisymmetric Landau Fermi parameter as a function of $\alpha_{gr}$ by using the renormalized Fermi velocity calculated
in Eq.(9). As it is clear from Eq.(14), this quantity is proportional to inverse renormalized Fermi density and its absolute value increases by increasing density. Consequently, the spin-susceptibility decreases by increasing impurity density as charge compressibility behaves~\cite{asgari_im}. In the inset, $F^a_0$ as a function of coupling constant is shown by using the on-shell renormalized Fermi velocity.

In Fig.~4 we show our theoretical predictions for the renormalized Fermi velocity of
doped graphene scaled by Fermi velocity of noninteracting clean system as a function of density $n$ in units of $10^{12} ~{\rm cm}^{-2}$ at $\alpha_{gr}=1$. In connection with experiments, the gate voltage $V$, and electron density are related through the gate capacitance per area, $C$ by $V=e n/C$. Therefore, the results of $v^*/v$ as a function of $n$ or $V$ could be directly tested by experiments. For definiteness we take $\Lambda=k_c/k_F$ to be such that $\pi (\Lambda k_F)^2=(2\pi)^2/{\cal A}_0$, where
${\cal A}_0=3\sqrt{3} a^2_0/2$ is the area of the unit cell in the
honeycomb lattice, with $a_0 \simeq 1.42$~\AA~the carbon-carbon
distance. With this choice $\Lambda\simeq{(g
n^{-1}\sqrt{3}/9.09)^{1/2}} \times 10^2$, where $n$ is the
electron density in units of $10^{12}~{\rm cm}^{-2}$. We have shown the related scattering rate as a function of electron charge density in the inset of figure.

As it is clear from Fig.~4 the renormalized Fermi velocity decreases by
increasing impurity density. By
increasing the interaction effects,
i.e., decreasing the density our theoretical effective Fermi velocity results decrease.
Furthermore, we have found that at low electron density regime by increasing the impurity density,
the renormalized Fermi velocity decreases sharply at $n_i=10^{10}~{\rm cm}^{-2}$ and reaches to a constant. Similar behavior occurs at larger impurity densities as well. Consequently, the impurity dependence of the renormalized Fermi density is quite nontrivial and nonmonotonic and this could be directly experimentally tested. Although, it is known that application of RPA in very low density in graphene is questionable, however it would be interesting
to explore this effect by going beyond of RPA or using a density functional type model calculation by incorporating correctly the physics of linear dispersion relation, chirality and the exchange-correlation energy. It would be useful to carry out further experimental work at low
electron density region to assess the role played
by correlation and impurities effects.

\section{Conclusion}

We have studied the quasiparticle properties of the chiral electron
graphene sheet within the Landau Fermi theory
based on the random phase approximation incorporating
the impurities in the system. Our approach is based on a
self-consistent calculation between impurity effect and
many-body electron-electron interaction. We have used a model
surface roughness potential together with the charged disorder
potential in the system. We find that there are significant differences
between quasiparticle properties of C2DES and conventional 2DEG due to disorder effects.
Our calculations of renormalized Fermi velocity
show a reduction about $5-15$ percent at experimental available density region.

We remark that in a very small density region, the system is highly
inhomogeneous and the effect of the impurities are very essential. A model going beyond
the RPA is necessary to account for increasing correlation effects
at low density. Our finding at low electron density would be verified by experiments.

\begin{acknowledgments}
We thank M. R. Ramezanali and M. Polini for useful discussions. A. \,Q. is supported by IPM grant.
\end{acknowledgments}

\end{document}